\documentclass[aps,prl,showpacs,showkeys,twocolumn,superscriptaddress,
              floats,floatfix,preprintnumbers,amsmath]{revtex4}

\usepackage{graphicx}
\usepackage{subfigure}
\usepackage{bm,ulem}
\usepackage[usenames,dvipsnames]{color}
\def \tauP {\tau^{\rm P}}
\def \Emax {\mathcal{E}_{\rm max}}

\def \grad {{\bm \nabla}}

\def \dive {{\bm \nabla}\cdot}
\def \lap {\nabla^2}

\newcommand{\deldel}[2]{\frac{\partial #1}{\partial #2}}

{}

\def \Rey  {\mbox{Re}}
\def \Ca  {\mbox{Ca}}
\def \Cb  {C_{\rm b}}

 \def \ac {a_{\rm c}}

\def \hini {h_{\rm ini}}
\def \tflow {t_{\rm flow}}
\def \mA {\mathcal{A}}
\def \mAz {\mathcal{A}_0}

\def \Gs {G_{\rm s}}
\def \Gb {G_{\rm b}}

\newcommand{\Eq}[1]{Equation~(\ref{#1})}

\newcommand{\fig}[1]{Fig.~\ref{#1}}
\newcommand{\subfig}[2]{Fig.~\ref{#1}(#2)}
\newcommand{\Fig}[1]{Figure~\ref{#1}}

\newcommand \bx {\bm{x}}

\def \rj {{\rm j}} 
\def \ff {\bm{f}}
\def \xj {\bm{x}_{\rm j}}
\def \lone {\lambda_{1}}
\def \ltwo {\lambda_{2}}
\def \Ione {\mathcal{I}_{1}}
\def \Itwo {\mathcal{I}_{2}}

\newcommand\bu{\ensuremath{\mathbf{u}}}

\newcommand\rmmax{\ensuremath{\mathrm{max} }}
\newcommand\tpmax{\ensuremath{\tau^{\mathrm{P}}_{\rmmax}}}

\begin{document}
%\title{A microfluidic device to sort capsules by deformability}
\title{A microfluidic device to sort capsules by deformability: A
  numerical }
\author{L. Zhu}
\affiliation{Linn\'e Flow Centre and SeRC (Swedish e-Science Research Centre), KTH Mechanics, SE-10044 Stockholm, Sweden}
\author{Cecilia Rorai}
\affiliation{Linn\'e Flow Centre and SeRC (Swedish e-Science Research Centre), KTH Mechanics, SE-10044 Stockholm, Sweden}
\affiliation{Nordita, KTH Royal Institute of Technology and Stockholm University,
Roslagstullsbacken 23, 10691 Stockholm, Sweden}
\author{Dhrubaditya Mitra}
\affiliation{Nordita, KTH Royal Institute of Technology and Stockholm University,
Roslagstullsbacken 23, 10691 Stockholm, Sweden}
\author{ Luca Brand}
\affiliation{Linn\'e Flow Centre and SeRC (Swedish e-Science Research Centre), KTH Mechanics, SE-10044 Stockholm, Sweden}
\begin{abstract}
Guided by extensive numerical simulations, we propose a microfluidic device that can sort elastic capsules
by their deformability.
The device consists of a duct embedded with a semi-cylindrical obstacle, and a diffuser which further enhances 
the sorting capability.
We demonstrate that the device can operate reasonably well under changes in the initial position of the
the capsule.
The efficiency of the device remains essentially unaltered under small
changes of the obstacle shape (from semi-circular to  semi-elliptic
cross-section).
Confinement along the direction perpendicular to the plane of the device increases
its efficiency.
This work is the first numerical study of cell sorting by a realistic microfluidic device.
\end{abstract}

\pacs{
}
\maketitle
%------------------
\section{Introduction}

One vitally important challenge in the field of biotechnology
is to design devices to sort cells by chemical and physical properties. 
These devices can be used for rapid medical diagnoses at the cellular level,
and screening to guard against deliberate contamination.~\cite{Pra03}
To quote a few specific examples,  such devices would be effective tools to 
(a) measure the altered deformability of Red Blood Cells (RBCs) 
e.g., due to malaria,~\cite{sur06} 
(b) sort  bacteria or yeast cells by their length, 
or  (c) extract  circulating tumour cells from blood of a cancer patient.~\cite{lim+hoo14}
An oft-used device in this category is a flow cytometer that can sort cells based on their
optical responses.~\cite{Pra03}

It is clear from the examples quoted above that  the physical properties of a cell, e.g., size,
shape, or deformability are important bio-markers; 
it is hence crucial to try to develop cell-sorting devices based on them.
Furthermore, these markers may even be preferable  to 
biochemical markers used in traditional medical diagnostics because they 
are label-free. 
A device to sort cells by biophysical markers  can be low cost, convenient to 
maintain, and characterized by shorter assay times and good throughput.~\cite{mao} 
As a further motivation, we note a remarkable use of biophysical markers in natural biological 
systems:  the spleen separates old and damaged RBCs
from healthy ones by passing them through slits between endothelial cells.  
Only  RBCs deformable enough  are recirculated back to the venous system, while ageing RBCs
are 
phagocytosed in the cord of the spleen red pulp.~\cite{spleen}

In recent times, several microfluidic devices have been fabricated to detect biophysical markers and to 
sort cells accordingly. \cite{hou10, bow11, holm11, hur11, gossett, lund, mao}
The challenge in this field lies in designing clever geometries that allow for an efficient sorting.
Mircofluidic devices possess the unique ability to sort cells by deformability because 
they operate by balancing the elastic stresses of the cell against the fluid stresses.
It then behoves us to try to  understand and model  flows carrying
suspended cells; let us elaborate on this point. 
Given the geometric configuration of a microfluidic device it is computationally straightforward to find out the
flow in the absence of cells.  
This is because the small size of microfluidic devices implies that the viscous effects
dominate over inertia and hence the solution to the flow
problem can be obtained by solving the linear Stokes equations. 
But as soon as a deformable object, e.g., a cell,  is introduced,  the mutual interaction between
the elastic stresses at the cell surface and the viscous fluid stresses
turns the problem into a formidable, nonlinear one. 

\begin{figure}[!ht]
 \hspace{1em}\includegraphics[width=0.92\columnwidth]{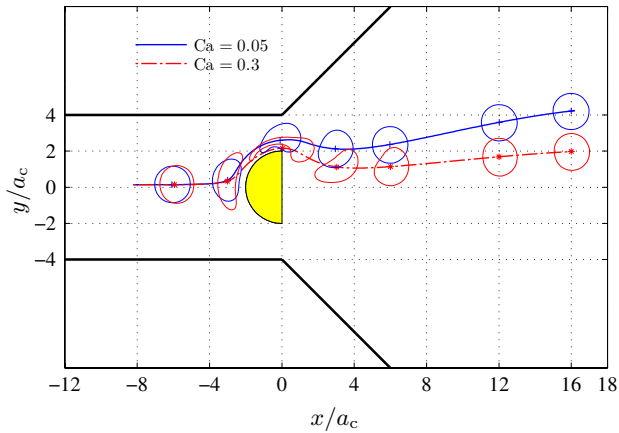}
 \put(-120,-10){$x/\ac$}
 \put(-235,67){\rotatebox{90}{ $y/\ac$}  }
 \caption{\label{fig:sketch}
A two-dimensional ($x$-$y$ plane) sketch of our computational domain with the paths of two 
capsules, a stiff one ($\Ca=0.05$, solid blue line)  and  a floppy one ($\Ca=0.3$, broken red
line), 
starting from the same initial position, at an offset $\hini=0.015\ac$ above the mid-plane
($y=0$). 
The flow is driven from left to right. 
The diffuser (the diverging duct) makes an angle of $45^{\circ}$ with the $x$ axis. 
All length scales are normalized by the equilibrium radius, $\ac$, of the capsule.
The extent of the device in the $z$ direction (perpendicular to the plane shown) is 
$H_z = 4\ac$.}
 \end{figure}

Over the last decade, numerical techniques and computational capabilities have developed 
hand-in-hand such that it is now possible to solve such microscale complex flows
in a computer.~\cite{freund_13_review,peng2013lipid} 
The time is now ripe to use simulations to complement and speed up the usual experimental
trial-and-error process required to perfect a microfluidic device.
As an example of such an exercise, in this paper we use extensive numerical simulations to
propose the design of a microfluidic device, sketched in \fig{fig:sketch}, that
can potentially sort cells by their deformability. 
%-------------
%-------------------

\section{Models}
Due to the molecular complexity of a cell and its sensitivity to the surrounding
environment~\cite{freund_13_review}, it is not possible at this moment to accurately account for the full material
property of the cellular structure. The membrane of RBCs has a
typical thickness of the order of nanometres~\cite{ufert2011cap_review_shear}, which is much
smaller than their size. We therefore mimic our cells by the capsule model that is a
droplet  enclosed by an infinitely thin hyperelastic sheet; it is
endowed with shear and bending
elastic resistance, representing the cellular spectrin network.
We further consider the sheet as isotropic; this assumption is
reasonable~\cite{freund_13_review} and has been widely
used~\cite{dbb11capreview,ufert2011cap_review_shear,freund_13_review}. 
This implies that the
local strain energy function $W(\Ione,\Itwo)$ is a function 
of only $\Ione = \lone^2+\ltwo^2-2$ and $\Itwo=\lone^2\ltwo^2-1$, which 
are the two invariants constructed from the two
principal components of the strain, $\lone$ and $\ltwo$. 
Among several possibilities we choose the oft-used neo-Hookean model~\cite{bar+wal+sal10} 
for which
\begin{equation}
W=\frac{\Gs}{2}\left[\Ione -1 + \frac{1}{\Itwo+1}\right],
\label{eq:nHookean}
\end{equation} 
where $\Gs$ is the isotropic shear modulus. 
Another commonly used alternative is the Skalak model, used e.g., in 
Refs.~\citenum{freund2011cellular,freund2013flow} for RBCs;
see also Ref.~\citenum{bar+dia+dhe02} for a comparison between several constitutive models.
We also employ a linear isotropic model for the bending moment,~\cite{zhao10}
with a bending modulus $\Gb= \Cb \ac^2 \Gs$, where $\ac$ is the radius of the capsule and 
$\Cb = 0.01$ is held constant in our simulations.
The choice of $\Cb$ is consistent with the available experimental data for
RBCs.~\cite{freund_13_review}
Finally we also assume that the fluid inside and outside the cell
has exactly the same density and viscosity;
and the equilibrium shape of the cell is a sphere of radius $\ac$. The model
is specifically designed for anucleate cells, since the presence of a stiff nucleus is not 
explicitly considered. In spite of this fact, the model is potentially applicable to nucleate
cells when the
elastic constants, e.g., $G_s$, are tuned to effectively account for the internal cellular 
structures.
To summarize, we solve for capsules with neo-Hookean membrane in flows.
Even stripped of all its biological context, this problem is interesting in its own right
for its potential applications to the fields of chemical engineering, bioengineering, and food
processing among others.

There are two dimensionless numbers in this problem, the Reynolds number, $\Rey \equiv \rho U \ac/\mu$,
and the capillary number, $\Ca  \equiv \mu U/\Gs$, where $U$ is the characteristic velocity, $\mu$ the dynamic viscosity
and $\rho$ the density. The capillary number expresses the ratio between the viscous and
elastic forces, which increases if the shear modulus decreases (softer capsules), or
equivalently, if the mean velocity increases (larger fluxes). Hence, our results can be interpreted as either
the dynamics of capsules with different deformabilities in the same flow, or that
of the same capsule under different flow conditions. The Reynolds number  is typically below $0.01$ in microfluidic devices by virtue of the 
small length scales involved. Hence we use the linear Stokes equations ($\Rey = 0$) to solve the flow. 

For the sake of completeness,  
we provide a short description of the numerical algorithm~\cite{freund_13_review,lai14,lailai_phd} we use. 
The surface of the capsule is  discretized into $N$ points; the
$\rj$-th point has the coordinate $\xj$.
In the spirit of immersed boundary methods;~\cite{mittal2005immersed} at the $\rj$-th point, 
a  force $\ff_{\rj}$ is exerted on the flow.
These forces are determined by the deformation of the capsule with respect to its 
equilibrium shape through an appropriate constitutive law; in this case the neo-Hookean model.
We use a spectral method~\cite{zhao10} to calculate $\ff_{\rj}$ given the
positions $\xj$.
The flow field can then be obtained by solving the Stokes problem,
with the forces $\ff_{\rj}$ added to the right hand side, i.e., 
\begin{eqnarray}
\label{eq:stokes}
-\grad p + \mu\lap\bu & = &- \sum_{\rj=1}^{N} \ff_{\rj}\delta\left(\bx -\xj
\right),  \\ 
\label{eq:divu}
\dive \bu &  =  &  0.
\end{eqnarray}
Here $p$ is the pressure, $\bu$ is the velocity, and $\delta$ denotes the 
Dirac delta function. 

Equations \ref{eq:stokes} and \ref{eq:divu} are solved by a hybrid Integral-Mesh
method.~\cite{kumar12,graham07_prl}  
In our implementation, the mesh-based part (responsible for the
long-range part of the Green's function) is calculated by the spectral-element solver
NEK5000~\cite{nek5000-web-page} which allows us to cope with non-trivial boundaries.
The short-range part is handled by standard boundary integral techniques. 
Once the Stokes problem is solved we know the velocity of the flow at every point 
including each point on the surface of the capsule, $\xj$; $\bu(\xj)$. 
The values of $\xj$ at the next time step are obtained by solving,

\begin{equation}
\frac{d \xj}{dt} = \bu(\xj)\/. 
\end{equation}
This implies that the $\rj$-th point on the surface of the cell moves with
the velocity, $\bu(\xj)$, i.e., a no-slip, non-penetrating boundary condition is satisfied
on the cell surface. 
With this algorithm, we are able to perform  high-fidelity simulations
of deformable capsules suspended in microfluidic flows with complex
domains.~\cite{lailai_phd}

\section{Results}
%-----------------------------------------
\begin{figure*}[!ht]
\centering
 \includegraphics[width=0.8\linewidth]{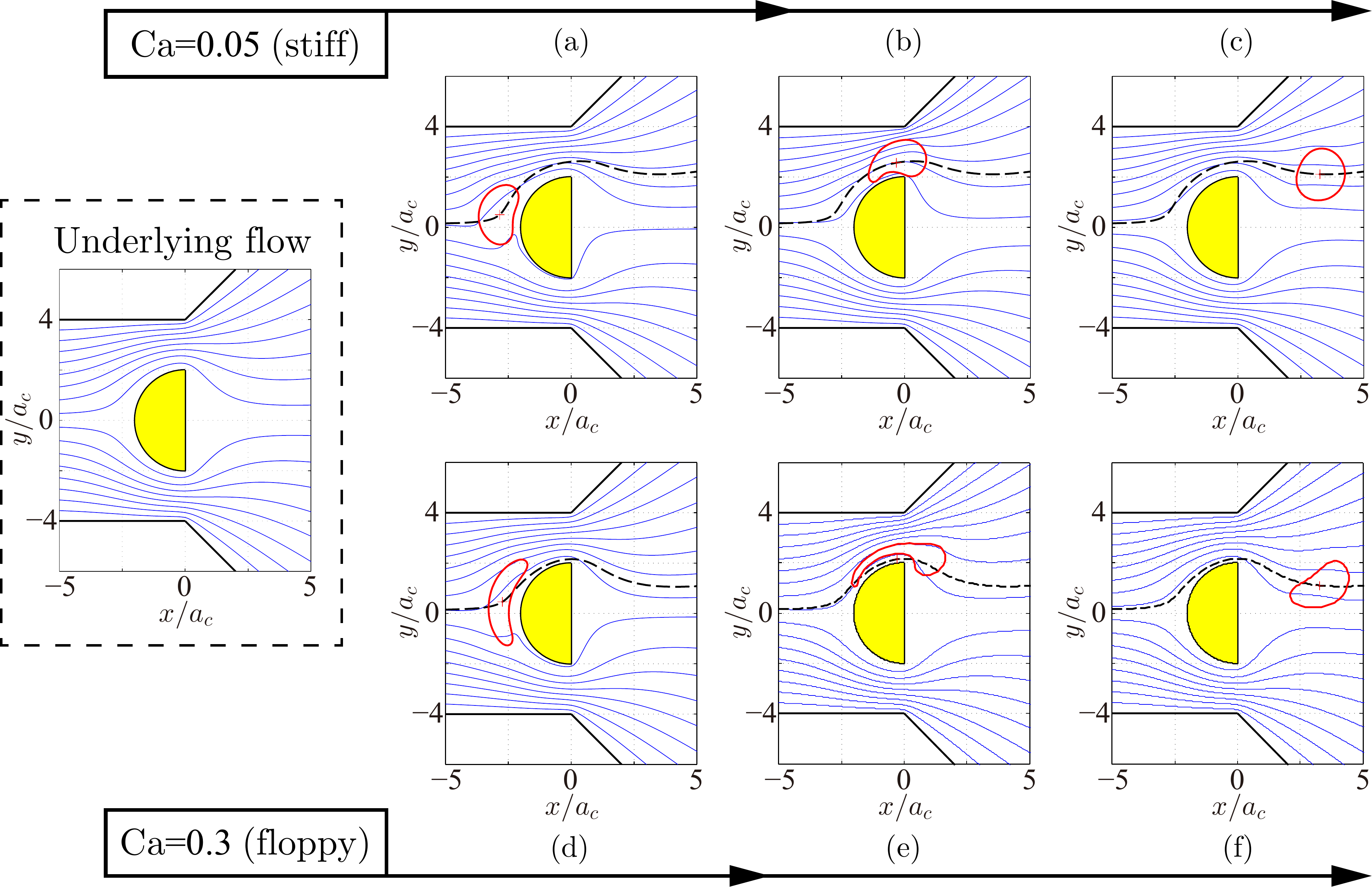}
\caption{
Two-dimensional profiles of two capsules at different time instants, plotted on the $z=0$
plane with flow streamlines. The plot in the dashed box displays the streamlines of the flow 
without capsules. The evolution of the stiff capsule $\Ca=0.05$ and the floppy one $\Ca=0.3$,
is illustrated in the top and bottom row respectively. The
centre of mass of the capsule is indicated
by the cross, and its trajectory  by the dashed line. Snapshots are taken for each
capsule at three time instants when: (a)/(d), the
capsule almost touches the obstacle; (b)/(e), it 
sits in the gap above the obstacle; (c)/(f), it is well past the obstacle.}
\label{fig:stream}
\end{figure*}
%-------------------
The device we propose is a rectangular duct attached to a diverging one (a diffuser),
as shown in \fig{fig:sketch}.
An obstacle which encompasses the entire depth
of the device ($z$ direction), is positioned at the junction of the duct and the diffuser symmetrically about the mid-plane ($y=0$).
A capsule, whose equilibrium shape is a sphere of radius $\ac$, is placed at the inlet.
Initially, the centre of the capsule is not on the mid-plane but is displaced (along the
$y$-direction) 
by an amount $\hini$.
The analytical flow profile~\cite{spiga1994symmetric} of a rectangular duct is maintained at the
inlet, extreme left in \fig{fig:sketch}, 
and zero-stress boundary conditions $-p\mathbf{I}+\mu\left(\nabla
\mathbf{u}+\left(\nabla \mathbf{u}\right)^{T}\right)=0$ are 
imposed at the outlet, extreme right in \fig{fig:sketch}.
The width and thickness of the duct (the extent of the $y$ and $z$ direction) is
$8\ac$ and $4\ac$, respectively; non-penetrating, no-slip boundary conditions are 
imposed on the wall of the device.

The functioning of the device is demonstrated by a series of images in \fig{fig:stream},
see also the animation showing the motion of the capsules through our device
at the location $http://www.youtube.com/watch?v=-1Nbq_QGpSs$.
The flow field in the absence of the capsule is plotted in \subfig{fig:stream}{a}. 
When the capsule is at the inlet, the flow is very similar to that without  
the capsule.
As the capsule approaches the obstacle, it slows down, deforms significantly (the
deformation depends on $\Ca$) and
substantially modifies the flow as shown in \subfig{fig:stream}{a}-({f}).
Due to the interaction between the elastic membrane and the viscous flow, the capsules follow
different paths depending on their deformability.
Two extreme cases are sketched in \fig{fig:sketch} which demonstrate that at
the outlet of the device two capsules with $\Ca=0.05$ (stiff) and $0.3$ (floppy) are clearly separated.  
This completes our primary objective, i.e., to demonstrate that our device can sort
capsules by deformability. 

\begin{figure}[!ht]
 \centering
 \includegraphics[width=0.30\columnwidth]{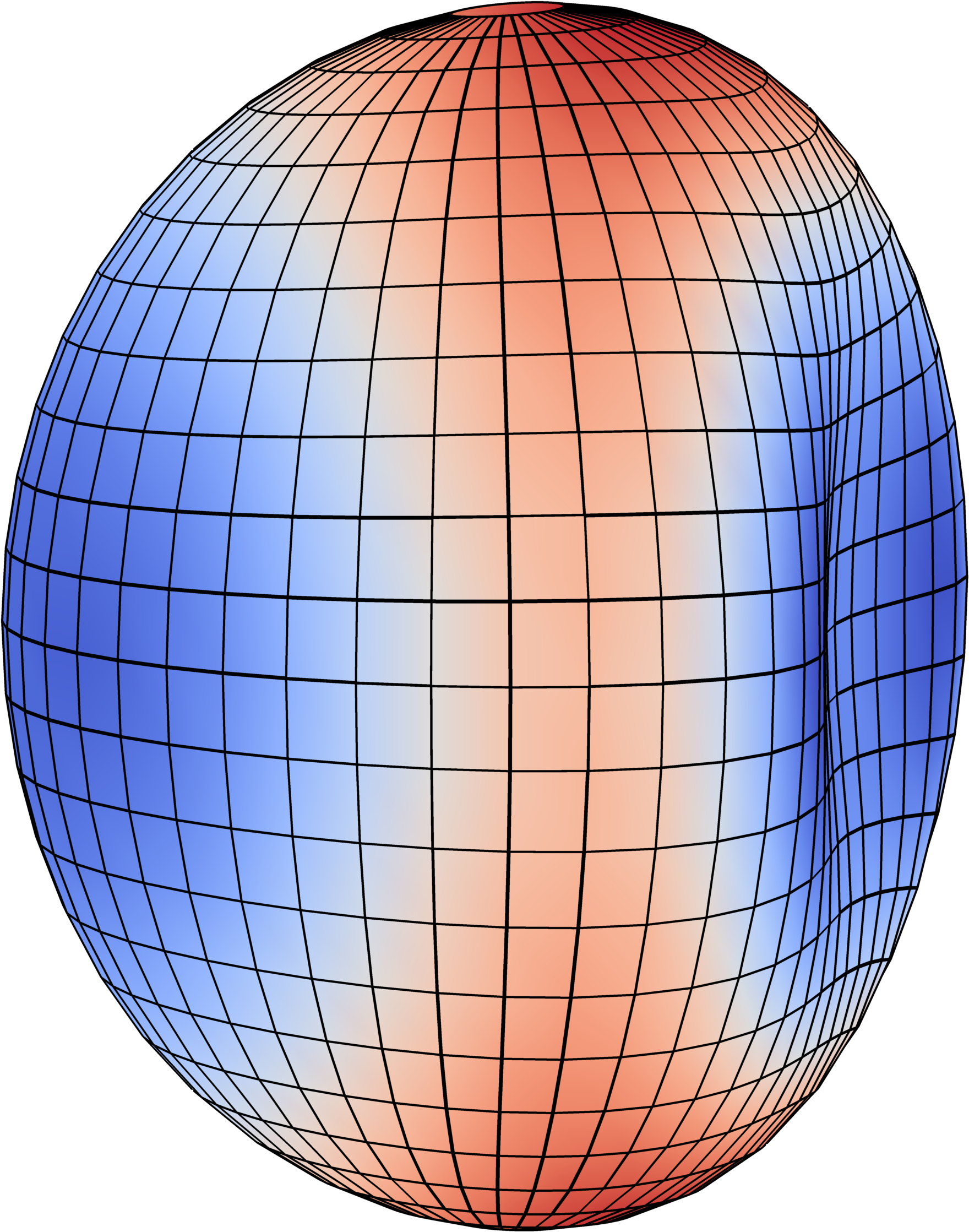}
 \put(-40,100){(A)}
 %\put(-100,50){(I)}
 \hspace{0.1cm}
 \includegraphics[width=0.30\columnwidth]{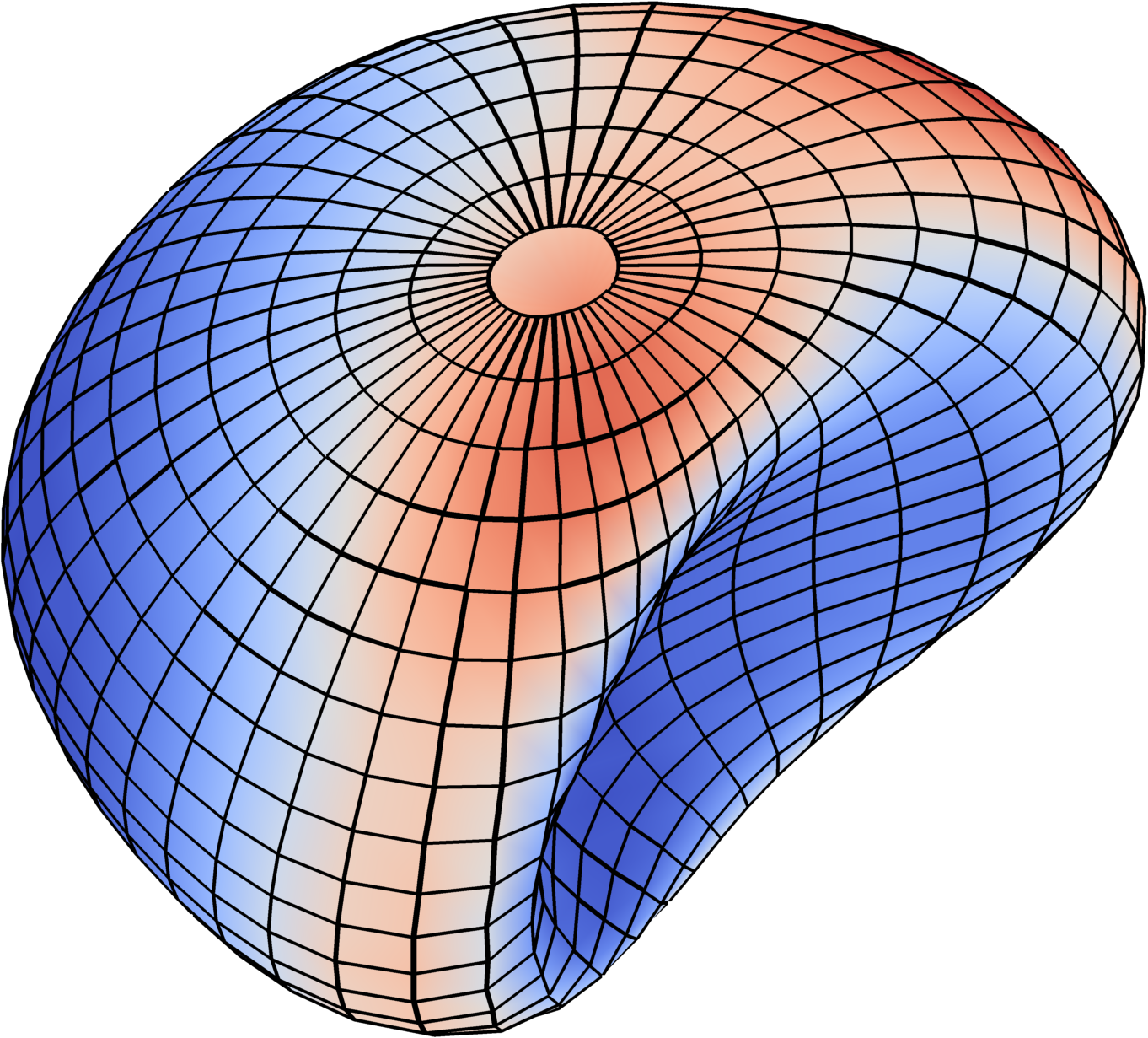}
 \put(-40,100){(B)}
 \hspace{0.1cm}
 \includegraphics[width=0.30\columnwidth]{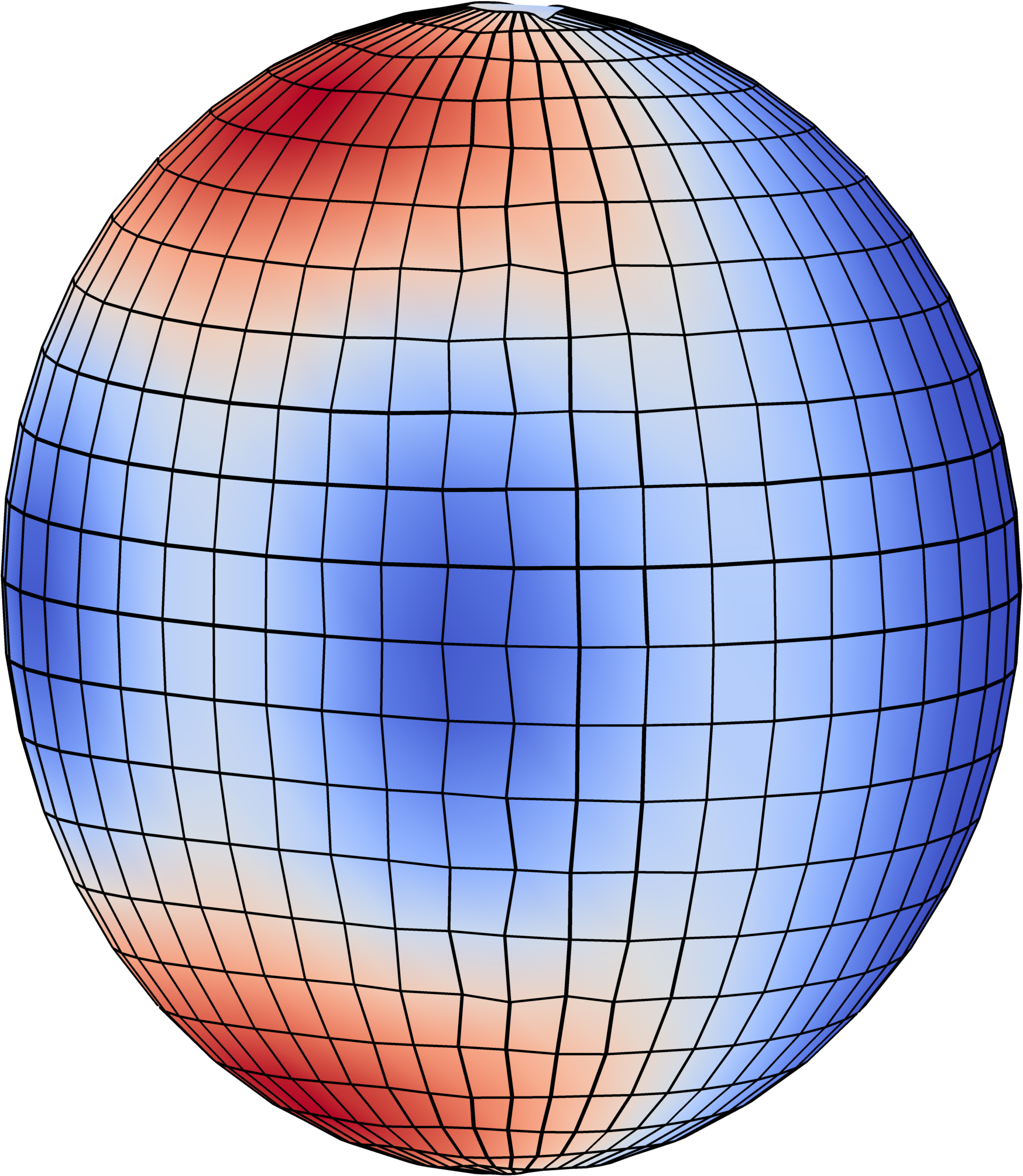} 
 \put(-40,100){(C)}\\
\vspace{0.1cm}
 \includegraphics[width=0.30\columnwidth]{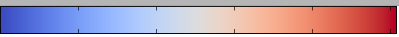} 
 \put(-80,-10){\small{$0$}}
 \put(-40,-10){\small{$\frac{1}{2}\Emax$}}
 \put(0,-10){\small{$\Emax$}} \\
\vspace{0.05cm}
\includegraphics[width=0.30\columnwidth]{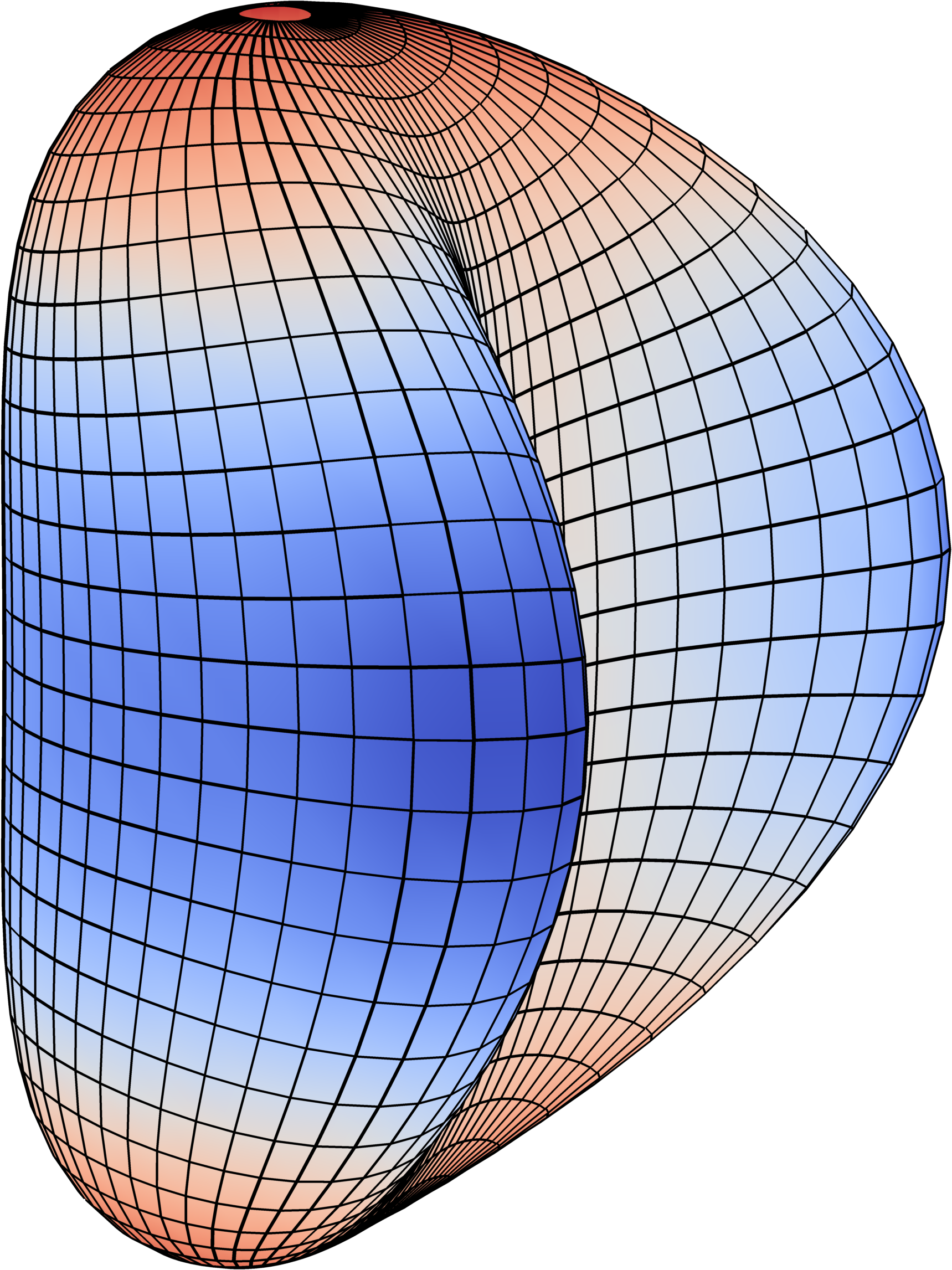}
 \hspace{0.1cm}
 \includegraphics[width=0.30\columnwidth]{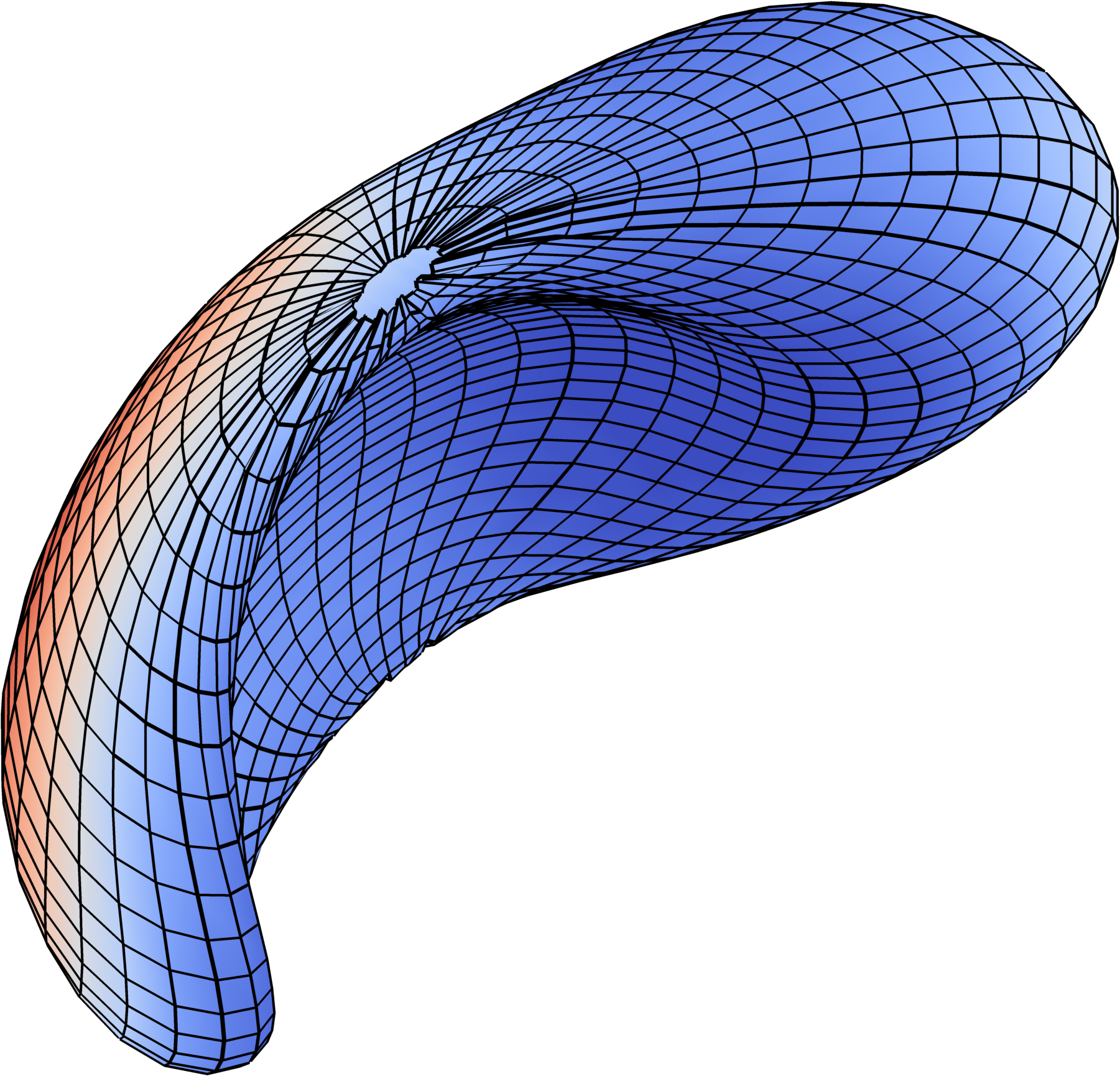} 
 \hspace{0.1cm}
 \includegraphics[width=0.25\columnwidth]{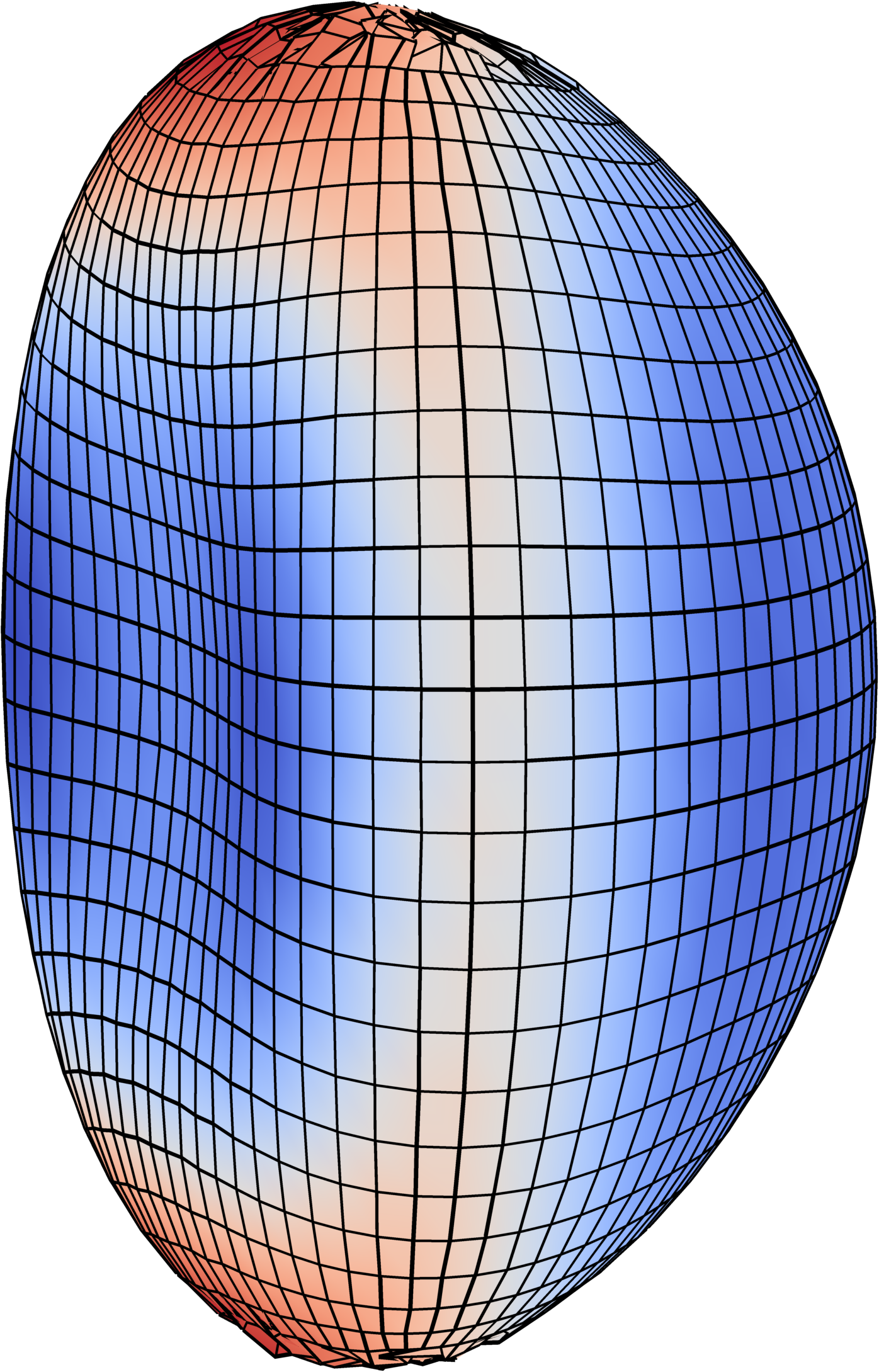}
 \caption{\label{fig:deform}
 Shapes of the capsule, with the computational grid sketched on the surface,
at different instances during its passage through the device. 
Top row: $\Ca = 0.05$. Bottom row: $\Ca = 0.3$. 
At a time when 
(A) the capsule almost touches the obstacle,
(B) the capsule sits in the gap between the obstacle and the duct-walls,
(C) the capsule has gone well past the obstacle. 
The pseudocolors show the variation of the local strain energy function 
non-dimensionalized by $\Gs\ac^2$; $\mathcal{E}\equiv W/(\Gs\ac^2)$.
Each plot is normalized by the maximum value of $\mathcal{E}$,
$\Emax$. 
In the top row, $\Ca=0.05$, $\Emax=0.04$ (A), $0.1$ (B), $0.004$ (C).
In the bottom row, $\Ca=0.3$, $\Emax=0.47$ (A), $3$ (B), $0.07$ (C).} 
\end{figure}

\begin{figure}[!ht]
\centering
\includegraphics[width=0.95\columnwidth]
{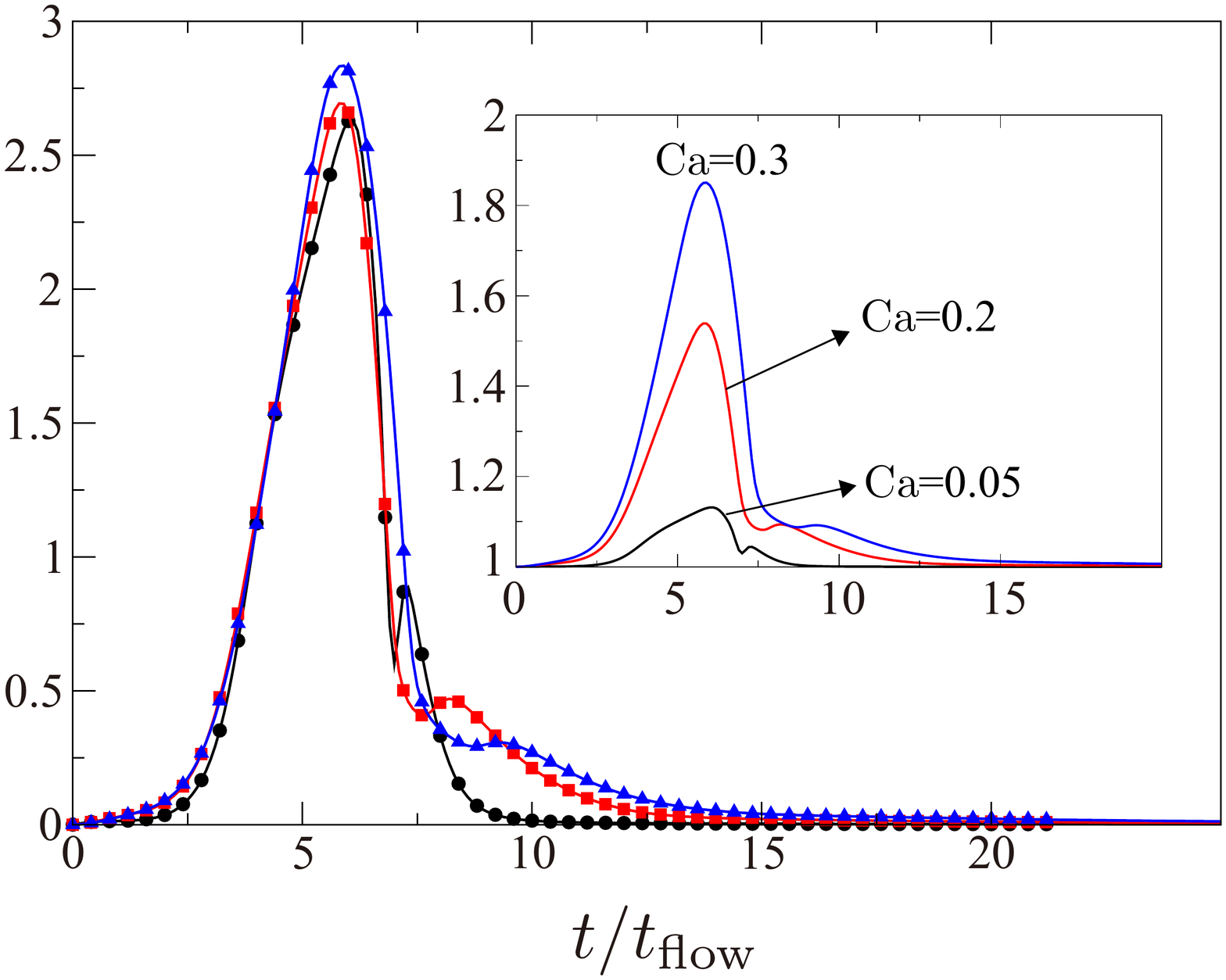}
\put(-235,55){\rotatebox{90}{ $\frac{1}{\Ca} \left( \mA/\mAz-1 \right) $}  }
\caption{\label{fig:area}
Fractional change in the surface area of the capsule, 
$\mA/\mAz-1$, for $\Ca=0.05$ (circle), $0.2$ (square), and $0.3$
(triangle) as a function of the nondimensionalized time  $t/\tflow$, where $\tflow \equiv \ac/U$.
When the vertical axis is scaled by $1/\Ca$ the three different curves collapse.
The inset shows the unscaled case. The area of the capsule in its equilibrium configuration is $\mAz =
4\pi\ac^2$. 
} 
\end{figure}

The deformed shapes of the capsules for $\Ca=0.05$ (stiff) and 
$\Ca=0.3$ (floppy) at different positions along their trajectories are shown in
\fig{fig:deform}.
When the capsule passes around the obstacle, it blocks the flow and
enhances the flow velocity on the opposite side of the obstacle; 
clearly, a stiffer (smaller $\Ca$) capsule produces a stronger blockage.
The deformation of the capsules are accompanied by large changes in their surface area as shown in the
inset of \fig{fig:area}.  Note that, the fractional change of the area of the capsules is 
roughly proportional to their capillary number as can be seen from the collapsed curves 
shown in \fig{fig:area}. 

%---------------------
%---------------------
%--------------------------------------- 
\begin{figure}[!ht]
\centering
\includegraphics[width=0.95\columnwidth]{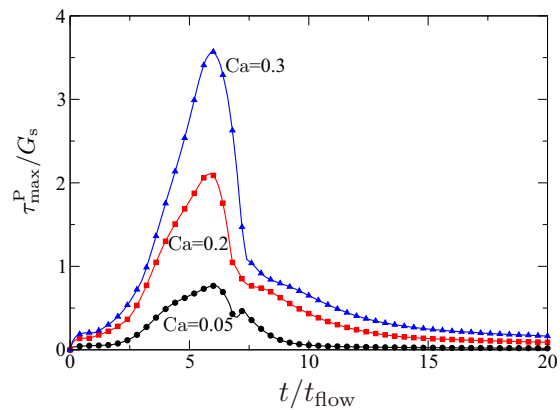}
\put(-228,75){\rotatebox{90}{ $\tpmax/\Gs$}  }
\caption{\label{fig:tension} The maximum value of the principal tension (\Eq{eq:tension}), 
nondimensionalized by $\Gs$, $\tpmax/\Gs$ as a function of time for 
$\Ca=0.05$ (black filled circles), $0.2$ (red squares), and $0.3$ (blue triangles).} 
\end{figure}

%------------------------------------------------------ 

The elastic stresses on the surface of the capsule are given by the
two principal tensions, $\tauP_1$ and $\tauP_2$, defined by:~\cite{skalak1973strain}
\begin{eqnarray}
\tauP_{1} &=& 2\frac{\lone }{\ltwo} 
        \left[ \deldel{W}{\Ione}+\ltwo^{2}\deldel{W}{\Itwo} \right], 
               \nonumber \\
  \tauP_{2} &=& 2\frac{\ltwo}{\lone}
     \left[ \deldel{W}{\Ione}+\lone^2\deldel{W}{\Itwo} \right].
\label{eq:tension}
\end{eqnarray}
The time evolution of the maximum stress, $\tpmax$, which is the
maximum value of $\tauP_1$ and $\tauP_2$ calculated over the
surface of the cell, is shown in \fig{fig:tension} for three
capillary numbers, $\Ca=0.05, 0.2$, and $0.3$.
The stress is maximum when the capsules pass through the gap
and increases with $\Ca$; for $\Ca=0.3$ it can be as large
as $3.5$ times $\Gs$. 
Clearly, too strong mechanical stresses can rupture capsules   
although rupturing depends not only on the maximum value of the
stress but also on how long it is applied. 
For example, RBCs at room temperature can even survive stresses up to
$5000$ Pascals for a very short time.~\cite{mus09}
The time evolution of the maximum stress in \fig{fig:tension} determines
the type of cells that can be sorted in this device.  

The basic working principle of this device is the following. 
To distinguish capsules by deformability we apply an external flow that forces
the capsules to pass through a narrow gap. 
The path of each capsule is then determined by the interaction between the viscous stresses and
the elastic stresses; a relative measure of these two is the capillary number, $\Ca$. 
For capsules with large $\Ca$ the viscous stresses dominate over the elastic ones, 
hence the trajectories of their centre-of-mass are close to the flow streamlines;  
these capsules deform far from their equilibrium shape. 
As an illustration, consider the limit $G_s\rightarrow \infty$, $\Ca=0$. 
In this limit, the membrane of the capsule does not resist deformation and its 
material points ($\xj$s) are advected by the flow as Lagrangian points.
Consequently their centre of mass follows the streamline of the underlying flow. 
In contrast, capsules with smaller $\Ca$ deform less and can alter the flow more,
hence their paths deviate further from the underlying flow; 
i.e., their centres are deflected further from the obstacle. 
The fact that a stiff capsule stays further from the obstacle than a soft one can also be understood in analogy
with
the collision of two capsules with different
deformability. It has been shown that stiff capsules are displaced more than soft ones in
heterogeneous
collisions.~\cite{kumar11,kumar12b} If the obstacle is regarded as a capsule with
infinitely
large stiffness, it follows, according to these previous observations, that the distance between the obstacle
and a moving
capsule decreases with its
deformability.

The basic working principle of our device is the same as that of 
the Deterministic Lateral Displacement (DLD) devices.~\cite{lund}
In all cases, the variation in the trajectories after passing around
a single obstacle is much smaller than the radius of the capsule, $\ac$.
Hence, by merely driving the capsules through a narrow gap (between the obstacle and the duct-walls)
it is not possible to generate large enough differences between their trajectories to separate
them.
The DLD device~\cite{lund} overcomes this limitation by using an array of obstacles
to generate an accumulative outcome.
We solve this problem by adding the diffuser where small displacements are magnified.
The Pinched Flow Fractionation devices~\cite{yamada2004pinched}, which sort particles by their
size, also use a diffuser. 
So far they have not been used to sort capsules by their deformability.
Note that the presence of obstacle is essential to achieve sorting. Removing the obstacle,
capsules with different deformability follow very close trajectories as shown by additional
simulations for capillary numbers $\Ca=0.05, 0.3$, and offset  $h_{ini}=0.15, 0.3$. 

Let us now discuss the typical length scales involved in designing this device. 
In the sketch shown in \fig{fig:sketch} all the length scales have been normalized 
by $\ac$.  The radius of the obstacle is $2\ac$. 
The gaps through which the capsules pass have a width of $2\ac$ too.
Although the exact values of these sizes are not crucial, they must  
be of the same order of the size of the cells. 

\begin{figure}[!ht]
\centering
\hspace{0em}\includegraphics[width=0.91\columnwidth]{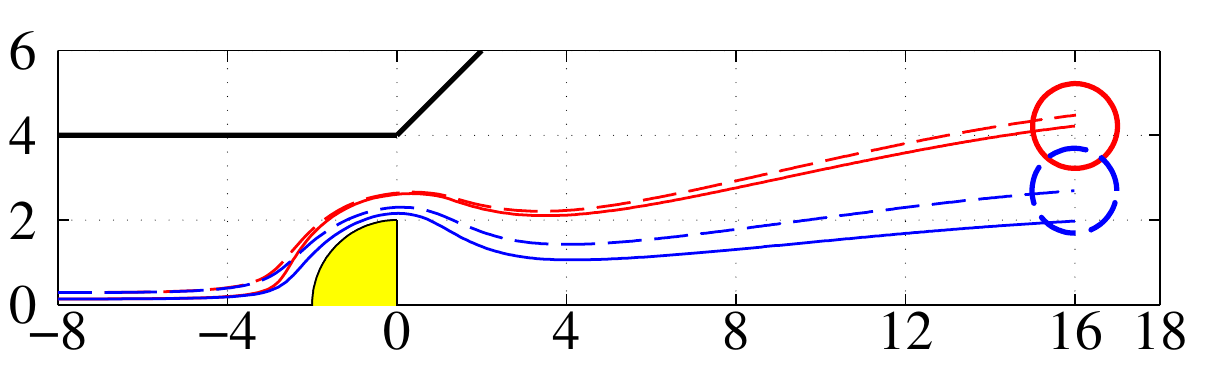} 
 \put(-120,-5){$x/\ac$}
 \put(-235,25){\rotatebox{90}{ $y/\ac$}  }
\put(-100,48){\Large{(a)}} \\
\hspace{-2em}\includegraphics[width=0.95\columnwidth]{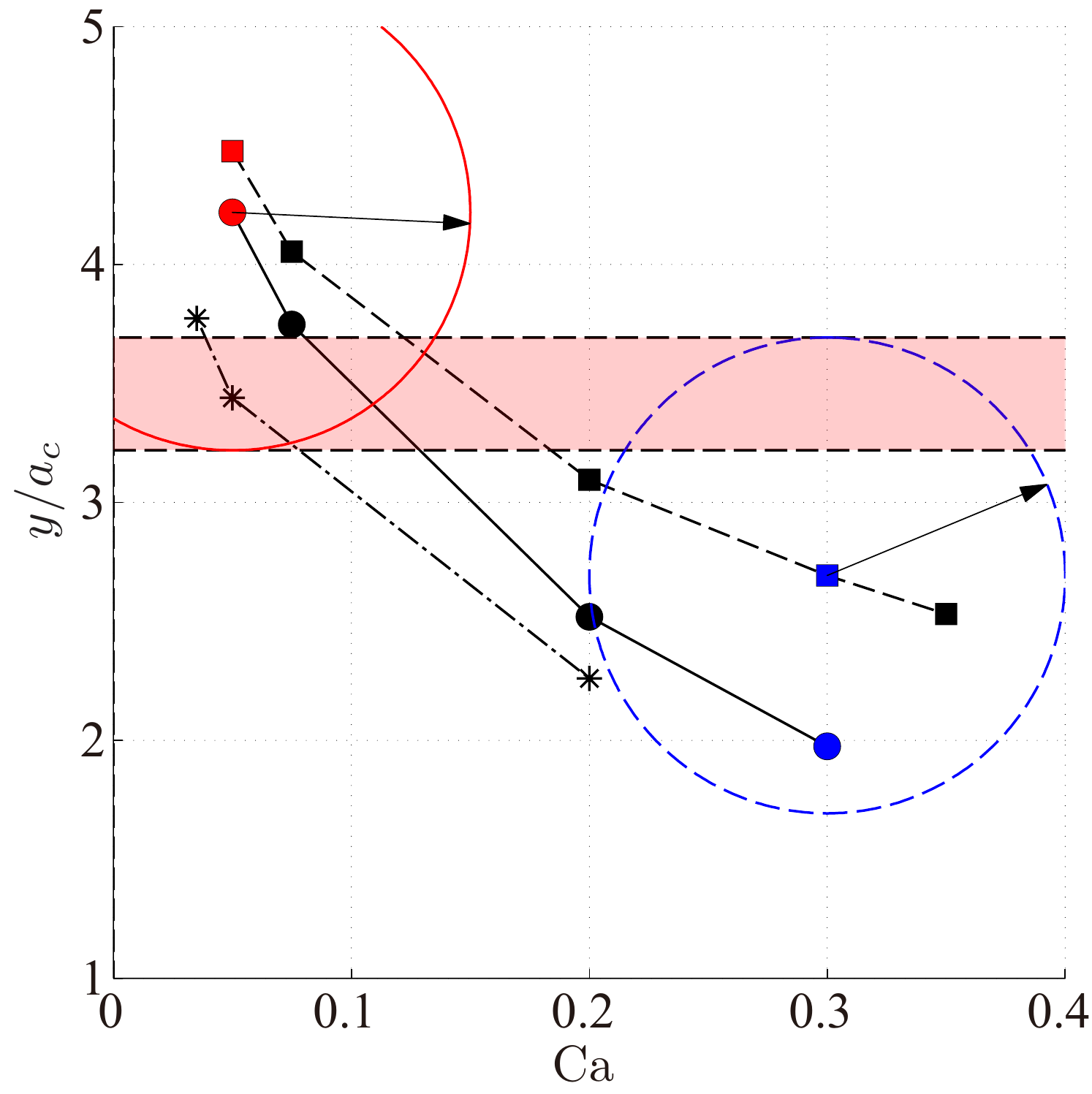}
\put(-90,180){\Large{(b)}} 
\caption{\label{fig:youtlet}
(a) Trajectories for $\Ca=0.05$ (stiff, red) and $\Ca=0.3$ (floppy, blue) for two initial offsets 
$\hini=0.15$ (solid) and $0.3$ (dashed). The profiles of the capsules, at the
outlet, clearly show an overlap. 
(b) The displacements of the centre of the capsules ($\Delta y$), above the mid-plane, measured
at
the outlet as a function of $\Ca$ for 
$\hini=0.15$ (solid) and $\hini=0.3$ (dashed).
The vertical axis is normalized by the equilibrium radius of the capsule, $\ac$. 
To calculate the overlap (the shaded region), 
for the pair of points $\Ca=0.05$, $\hini=0.3$ (red filled circle) and 
$\Ca=0.3$, $\hini=0.15$ (blue filled square), 
we draw circles of unit radius with the
centres of the capsules as their respective origins. 
The overlap for any pair of points can be calculated in a similar manner. 
The dashed-dotted line with the symbol asterisk ($\ast$) corresponds to the "unconfined" case
where we use periodic boundary conditions along the $z$ direction.}
 \end{figure}

Symmetry dictates that,  if initially the capsule is placed exactly on
the mid-plane it would get stuck in front of the obstacle,
in the absence of thermal fluctuations. Estimates show that thermal fluctuations can be ignored in this
problem.~\cite{freund_13_review}
Even without Brownian fluctuations, in an experimental realization, 
it would certainly be impossible that all the capsules are placed near the inlet with a precise
initial offset, $\hini$, from the mid-plane.
How do small changes in $\hini$ affect the sorting capability of our device?
To answer this question we run simulations for a range of values of $\Ca$, each  
with several different values of $\hini$.
Let us concentrate on the stiffest capsule, $\Ca = 0.05$, and the most floppy one, $\Ca = 0.3$.
If they are released from the same initial offset ($\hini$) then at the outlet the vertical 
displacement between their centres is larger than their diameter ($2\ac$), i.e., 
they are clearly separated, as shown in \fig{fig:sketch}.
Such a clear separation is not achieved if the capsules are released from 
different $\hini$s. For example, the trajectories for the
two cases, $\hini=0.3, \Ca=0.3$  and $\hini=0.15, \Ca=0.05$, are such that
at the outlet the exit regions for the two cases overlap, see \fig{fig:youtlet}(a).
To estimate this overlap, we plot, in \fig{fig:youtlet}(b), 
the displacements of the centre of the capsules from the mid-plane (at the outlet), $\Delta
y$, 
as a function of $\Ca$, for $\hini=0.15$ and $0.3$. 
The overlap between the capsules in \fig{fig:youtlet}(a) corresponds to the
shaded region in \fig{fig:youtlet}(b). 
\Fig{fig:youtlet}(b) can be similarly used to calculate the overlap between any 
two points in the figure.  
\Fig{fig:youtlet}(b) clearly shows that the paths of floppy capsules 
(larger $\Ca$) are more affected by the change in $\hini$. 
Note that, the overlap is quite small compared to $\ac$ and can be reduced
by using a longer diffuser. 

%-------------------

To further understand how crucially the performance of this device depends on its design, we
remove its spanwise confinement and impose periodic boundary conditions along the $z$
direction; the results are plotted in \fig{fig:youtlet} with the label ``unconfined''.
As expected, the absence of the spanwise confinement implies that for the same value of
$\Ca$, the vertical displacement decreases, in other words, the sorting capability of the
device becomes weaker.
Note that this comparison is made with the mean velocity of the underlying flow being held fixed. 
Next, instead of a perfect semi-cylindrical post, we test two more obstacles with
semi-elliptic cross sections, made by stretching the semi-cylinder in the $x$
direction by a factor of $\xi=2$ and $\xi=2/3$. Their cross sections are then two 
semi-ellipses with a major axis along the $x$ and $y$ direction, respectively.
We find that the displacements between capsules with different deformability are essentially
not altered by this geometric change, there is however a minor improvement in the sorting
capability when the major axis is oriented along the $x$ direction ($\xi=2$).

Finally, to understand the effects of the non-spherical shape of real biological cells, e.g.,
the
bi-concave shape of RBCs, we investigate 
an initially oblate capsule. Its axes along the revolution axis 
and orthogonal directions are denoted by $\ac^{\rm r}$ and $\ac^{\rm o}$ respectively and
their ratio $\ac^{\rm r}/\ac^{\rm o}=1/1.2$; the volume is taken to be the same as the
spherical capsules considered above. Two cells with $\Ca=0.05$ and $\Ca=0.3$ are simulated, with
revolution axes initially oriented in the $y$ direction. The displacement
between the two cells at the outlet of the device decreases only by about $0.5\%$ when compared to that between
spherical capsules. Note that RBCs are oblates with an aspect ratio of around $3$,  
and they can  tumble in shear flows~\cite{ufert2011cap_review_shear}. In such cases, the performance of our device
may deteriorate as the cell motion is not only affected by the deformability but also by the orientation of cells when
approaching the obstacle.

\section{Conclusions}
We model cells as fluid-filled capsules enclosed by neo-Hookean membranes characterized by two
elastic moduli, the shear modulus $\Gs$ and the bending modulus $\Gb$; the ratio between the
two is held constant.
Depending on the type of targeted cells, this model needs to be calibrated with 
the elastic measurements performed on the particular cells.
For human RBCs, different methods, e.g., the measurements by micro-pipette~\cite{Liu07}
or that by optical tweezers~\cite{henon99, lenormand01} give slightly different
values of $\Gs$. Diseases, e.g., sickle cell anaemia, can change this elastic coefficient by a factor of two to
three,~\cite{Lei12} and malaria can change the same elastic coefficient by a factor about
ten.~\cite{sur06} This is consistent with the range of capillary number
studied here. 
If we take a representative value of $\Gs \approx 2.5 \mu {\rm N} /{\rm m}$~\cite{sur06}, 
and use water as our fluid, then a typical flow rate of 
$0.1\mu$litre-per-minute gives $\Ca \approx
0.06$ which is well within the operating range of our device. We do not know the performance of the new
device as the capillary number is outside the range we have studied; further work will be
undertaken accordingly.
As the sorting behaviour depends on $\Ca$ and not on $\Gs$ alone,  
the same device can be used in a different range of $\Gs$ values by merely changing the flow
rate. A future direction is thus to investigate the sensitivity of the proposed mechanisms to inertial effects as the
flow rate is much higher.

To compare against other cell-sorting devices, the throughput of this device will be of the
same order as that of the optical flow cytometers~\cite{Pra03} because 
we can only allow one cell at a time to pass through the device. 
Margination devices~\cite{hou10} have a higher throughput as they operate on very
many cells simultaneously, but the accuracy of the device we propose is expected to be
significantly higher.
We further note that optical flow cytometers can be modified to sort by deformability 
too by adding obstacles in the path of the cells and by optical recognition of their
deformation. 
Due to the relatively simple model of the cell membrane used, and 
the fact that possible surface interactions between the cells and the walls of the device
have been ignored, we do believe that further refinements of the suggested devices are 
possible.
This work is the first numerical study of cell sorting in a realistic microfluidic
device and shows how accurate 
simulations may guide the initial stages of the design of new devices. 
We hope that our work will inspire the experimental
realization of devices based on the mechanism presented. 

\section{Acknowledgements}
The work has been financially supported by grants from the Swedish Research Council (to CR and DM),
the G\"{o}ran Gustafsson foundation (CR) and the European Research Council, Advanced Grant,
AstroDyn (DM). 
We thanks A. Brandenburg, R. Eichhorn and particularly D. Paul for many helpful discussions on
possible experimental realization of the device.
DM thanks IIT-B for hospitality, where a part of this work was carried out. 
Computer time provided by SNIC (Swedish National Infrastructure for Computing)
is gratefully acknowledged. We thank the anonymous referee for pointing out the possible limitations of our device
in sorting highly non-spherical cells.

%------------------

%-------------------
\bibliography{capsule.bib}
\end{document}